\pdfoutput=1
\documentclass{JINST}

\title{The High Resolution X-Ray Imaging Detector Planes for the MIRAX Mission}

\author{B{\'a}rbara H. G. Rodrigues$^{a,b}$,  Jonathan E. Grindlay$^b$, Branden Allen$^b$, Jaesub Hong$^b$, Scott Barthelmy$^c$, Jo{\~a}o Braga$^a$, Flavio D'Amico$^a$, Richard E. Rothschild{$^d$}\\
\llap{$^a$}Instituto Nacional de Pesquisas Espaciais (INPE),\\
  Av. dos Astronautas, 1758, S{\~a}o Jos{\'e} dos Campos, SP, 12227-010, Brazil\\
\llap{$^b$}Harvard-Smithsonian Center for Astrophysics (CfA)\\
  60 Garden St., Cambridge, MA, 02138,USA\\
\llap{$^c$} NASA Goddard Space Flight Center\\
  Greenbelt, MD, 20771, USA\\
\llap{$^d$} University of California-San Diego (UCSD)\\  
   9500 Gilman Drive, La Jolla, CA,  92093, USA\\ 
  E-mail: \email{barbara@das.inpe.br}}

\abstract{The {\it MIRAX}\/ X-ray observatory, the first Brazilian-led astrophysics space mission, is designed to perform an unprecedented wide-field, wide-band hard X-ray (5-200\thinspace keV) survey of Galactic X-ray transient sources. In the current configuration, {\it MIRAX} will carry a set of four coded-mask telescopes with high spatial resolution Cadmium Zinc Telluride (CZT) detector planes, each one consisting of an array of 64 closely tiled CZT pixelated detectors. Taken together, the four telescopes will have a total detection area of 959\thinspace cm$^{2}$, a large field of view ($60^{\circ} \times 60^{\circ}$\thinspace FWHM), high angular resolution for this energy range (6\thinspace arcmin) and very good spectral resolution ($\sim2$\thinspace keV @ 60\thinspace keV). A stratospheric balloon-borne prototype of one of the MIRAX telescopes has been developed, tested and flown by the Harvard-Smithsonian Center for Astrophysics (CfA) as part of the {\it ProtoEXIST} program. In this paper we show results of validation and calibration tests with individual CZT detectors of the {\it ProtoEXIST} second generation experiment (P2). Each one of 64 detector units of the P2 detector plane consists of an ASIC, developed by Caltech for the NuSTAR telescope, hybridized to a CZT crystal with $0.6$\thinspace mm pixel size. The performance of each detector was evaluated using radioactive sources in the laboratory. The calibration results show that the P2 detectors have average energy resolution of $\sim2.1$\thinspace keV\thinspace @\thinspace 60\thinspace keV and $\sim2.3$\thinspace keV\thinspace @\thinspace 122\thinspace keV. P2 was also successfully tested on near-space environment on a balloon flight, demonstrating the detector unit readiness for integration on a space mission telescope, as well as satisfying all MIRAX mission requirements.}


\begin{document}

\section{Introduction}
	
The Monitor e Imageador de Raios X (MIRAX) ~\cite{Braga04, Damico06} is the first Brazilian-led space observatory dedicated to astrophysics. The mission will perform an unprecedented high-cadence, wide-field (60$^{\circ} \times 60^{\circ}$), wide-band (5 to 200\thinspace keV) deep monitoring of Galactic X-ray transient sources. In its latest design, MIRAX will carry a set of 4 identical Hard X-Ray Imagers (HXI). The X-ray imaging detectors for these HXIs are currently being developed at the Harvard-Smithsonian Center for Astrophysics (CfA) in close cooperation with the National Institute for Space Research (INPE) in Brazil, which leads the MIRAX development. However, it is important to note that current funding constraints may change the design of the experiment in a way that a smaller number of HXIs will actually be onboard, together with a different CZT-based hard X-ray imaging camera provided by INPE, Brazil. Other important international partners of the mission include the University of California San Diego, which will provide the onboard tagged calibration source assembly device, and MIT and the University of Erlangen-Nuremberg in Germany, which will be co-responsible for the development of tools and pipelines for data reduction, analysis and storage at the MIRAX mission centers. MIRAX is scheduled to be launched in near-equatorial low Earth orbit in 2018 as part of the payload of the Lattes satellite. 

The MIRAX main instruments will be a set of coded-mask imagers, each including a close-tiled array of Cadmium Zinc Telluride (CZT) detectors. CZT is a direct bandgap semiconductor which is very attractive to applications in X-ray instruments due to their high average atomic number, providing high photoelectric absorption efficiency especially for hard X rays, and suitable band gap ($\sim 1.5$\thinspace eV), which allows for operation in direct conversion mode at room temperature. These advantages led to the development of CZT and CdTe hard X-ray detector systems which are currently operating on board three hard X-ray space missions: the ISGRI instrument on the INTEGRAL mission~\cite{Ubertini03}, the Burst Alert Telescope (BAT) on-board Swift ~\cite{Gehrels04}, and most recently the NuSTAR ~\cite{Harrison10} satellite. Both Swift and INTEGRAL carry coded mask telescopes with a large number of planar detectors which have limited spectral resolution (3.3\thinspace keV for Swift/BAT and 5.5\thinspace keV for ISGRI/INTEGRAL, both at 60\thinspace keV). NuSTAR is the first focusing hard X-ray (up to 79\thinspace keV) telescope mission in orbit, and it utilizes hybrid pixelated CZT detectors in its focal planes. In these detectors, in which the anode is divided into small unit pixels for high spatial resolution, advanced technology developed recently enables small leakeage current and low electronics noise, providing good spectral resolution ($\sim$ a few\thinspace keV\thinspace @\thinspace 60\thinspace keV). 

The recently developed and launched missions have given extraordinary contributions for the development of low cost CZT cystals and continue to provide a better understanding of the high energy astrophysics phenomena, including the discovery of many new sources. The next challange is to develop a wide-field sky-monitor which can combine both good angular and spectral resolutions for the investigation of the high energy X-rays transients in both soft and high energy bands of the spectrum. Currently, among the high energy missions in orbit which are capable to perform sky monitoring, INTEGRAL, Swift and Fermi~\cite{Atwood09} are not sensitive to energies $\lesssim 15$ keV, and MAXI ~\cite{Atswoka09} operates only in the 0.5-30\thinspace keV range. In addition, these telescopes have significatively poorer angular resolutions ($12^{\prime}$ (ISGRI), $22^{\prime}$ (BAT), several degrees (GBM), and $1.5^{\circ}$ (MAXI)) than MIRAX ($6^{\prime}$).  

In order to leverage the development of the architecture of the next generation of pixelated CZT hard X-ray survey telescopes, CfA is leading the {\it ProtoEXIST} program\footnote{http://hea-www.harvard.edu/ProtoEXIST/} (~\cite{Hong06, Hong09, Allen10, Hong11}) which consists of the development, assembly and tests of a series of moderate area ($256$ cm$^{2}$), fine pixel, coded-mask aperture telescope experiments. The second unit telescope, the {\it ProtoEXIST2} (P2), has been developed and successfully flown in a balloon-borne experiment from Fort Sumner, NM, in early October 2012 ~\cite{Hong13}, together with the first prototype telescope developed by the {\it ProtoEXIST} program, P1. P2 is the basic detection module for the coded masks telescopes of MIRAX. 

In this paper we present a brief description of the MIRAX mission, as well as the basic architecture of the pixelated CZT Detector Crystal Units (DCUs) which compose the P2 and MIRAX detector planes. We also show first results of characterization and calibration of the individual DCUs during the development phase. Finally we discuss the performance obtained with single DCUs using $^{241}$Am and $^{57}$Co radioactive sources.	

\section{The MIRAX mission}

MIRAX is part of the scientific payload of the Brazilian Lattes Satellite - a scientific space mission approved by the Brazilian Space Agency (Ag{\^e}ncia Espacial Brasileira (AEB)) - scheduled to be launched in 2018, carrying two scientific payloads - MIRAX and EQUARS (Equatorial Atmosphere Research Satellite) - in a multi-mission platform developed by INPE. The MIRAX mission, in its latest configuration, is designed to perform a hard X-ray ($5-200 $ keV) survey of more than half of the sky, concentrating the exposure time on the Galactic plane, with $6^{\prime}$ angular resolution, and with sub-millisecond time resolution. These characteristics are suitable for the investigation of all sorts of variable and transient phenomena in high energy astrophysics.

\begin{figure}[!htb]
  \begin{center}
    \includegraphics[width=\textwidth]{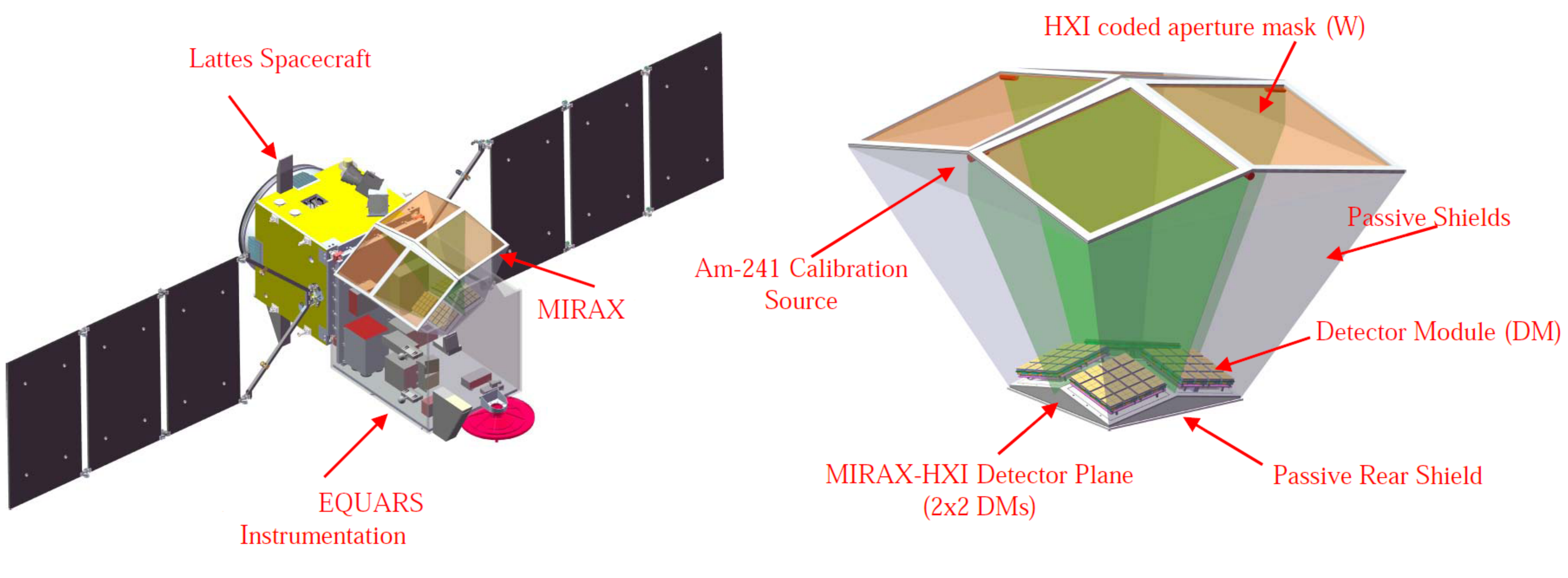}
    \caption{The Lattes satellite (left) is the first Brazilian scientific mission designed by INPE. It is scheduled to be launched in 2018, carrying two scientific payloads - MIRAX and EQUARS. The current design of MIRAX is the set of four coded mask hard X-rays (5-200 keV) telescopes (right), with combined $60^{\circ}\times60^{\circ}$ FWHM field-of-view (FoV).}
    \label{lattes_HXI}
  \end{center}
\end{figure}

The hard X-ray band is particularly important to study accretion onto black holes and neutron stars, including non-thermal processes of emission such as the ones occuring on jets and other hot relativistic astrophysical plasmas. MIRAX main goals include the study, with unprecedented depth and time coverage, of a large sample of transient and variable phenomena on accreting neutron stars and black holes, as well as Active Galactic Nuclei (AGNs) and Blazars and both short and long GRBs. This will be achieved by the combined operation of a set of coded-aperture telescopes that will operate in scanning mode in a near-Equatorial circular LEO for a lifetime of $3+$ years. Figure ~\ref{lattes_HXI} illustrates the Lattes Satellite, the current design of the telescopes and their components. 

The MIRAX detector development is a cooperative effort led by CfA (USA) together with NASA's GSFC, Caltech, INPE, and UCSD. The MIRAX detectors for the HXIs consist of a close-tiled array of $19.9 \times 19.9 \times 5$\thinspace mm CZT detectors, each with $32 \times 32$ pixels ($0.6$\thinspace mm pitch, same as NuSTAR), providing $959$\thinspace cm$^{2}$ imaging detection area (at 10\thinspace keV, total for the 4 units). The energy resolution obtained with these detectors is expected to be $\sim2$\thinspace keV (FWHM) at $60$\thinspace keV. A graded passive side shielding composed of Pb/Sn/Cu/Al layers will enclose the telescopes, except for the aperture opening. Four Tagged Calibration Source Assembly (TCSA) provide 60\thinspace keV photons from $^{241}$Am for monitoring the gain and offset of all CZT detectors continuously throughout the mission. For sources in the $50^{\circ}\times 50^{\circ}$ fully-coded field of view the sensitivity of MIRAX will approach that of Swift/BAT in the $15-150$\thinspace keV range, whereas the low threshold will enable $\sim70$\thinspace mCrab\footnote{Crab is a standard photometric unit defined as the intensity of X-rays emitted from the Crab Nebula and Pulsar at a given photon energy up to 30 kiloelectronvolts. In the 2-10 keV energy range,1 Crab is equal to 2.4$\times10^{−8}$\thinspace erg\thinspace cm$^{−2}$\thinspace s$^{−1}$\thinspace=15 keV\thinspace cm$^{−2}$\thinspace s$^{−1}$. The Crab flux is often used for calibration of X-ray telescopes.} sensitivity on time scales of $100$\thinspace s at $5-15$\thinspace keV energies, inaccessible to Swift/BAT and INTEGRAL. The 96\thinspace min (the orbital period) cadence of the MIRAX detections will enable simultaneous and follow-up observations in other wavelengths. In particular, there is a very interesting possibility of detecting short gamma-ray bursts (SGRB) in an epoch in which the Advanced-Ligo gravitational wave detector ~\cite{Abbott04} will be operational. A simultaneous SGRB and a gravitational wave signal will be of paramount importance both to gravitational wave physics and to the merging neutron star/neutron star model for short GRBs.

The pixelated CZT detectors that have been developed at CfA for P2 balloon-borne experiment satisfy all the requirements for the MIRAX mission. The detectors performance have been extensively tested with laboratory experiments and in a series of successful high-altitude balloon flights, demonstrating their readiness for space applications.

\section{The {\it ProtoEXIST2} telescope}

\begin{figure}[!htb]
  \begin{center}
    \includegraphics[width=0.6\textwidth]{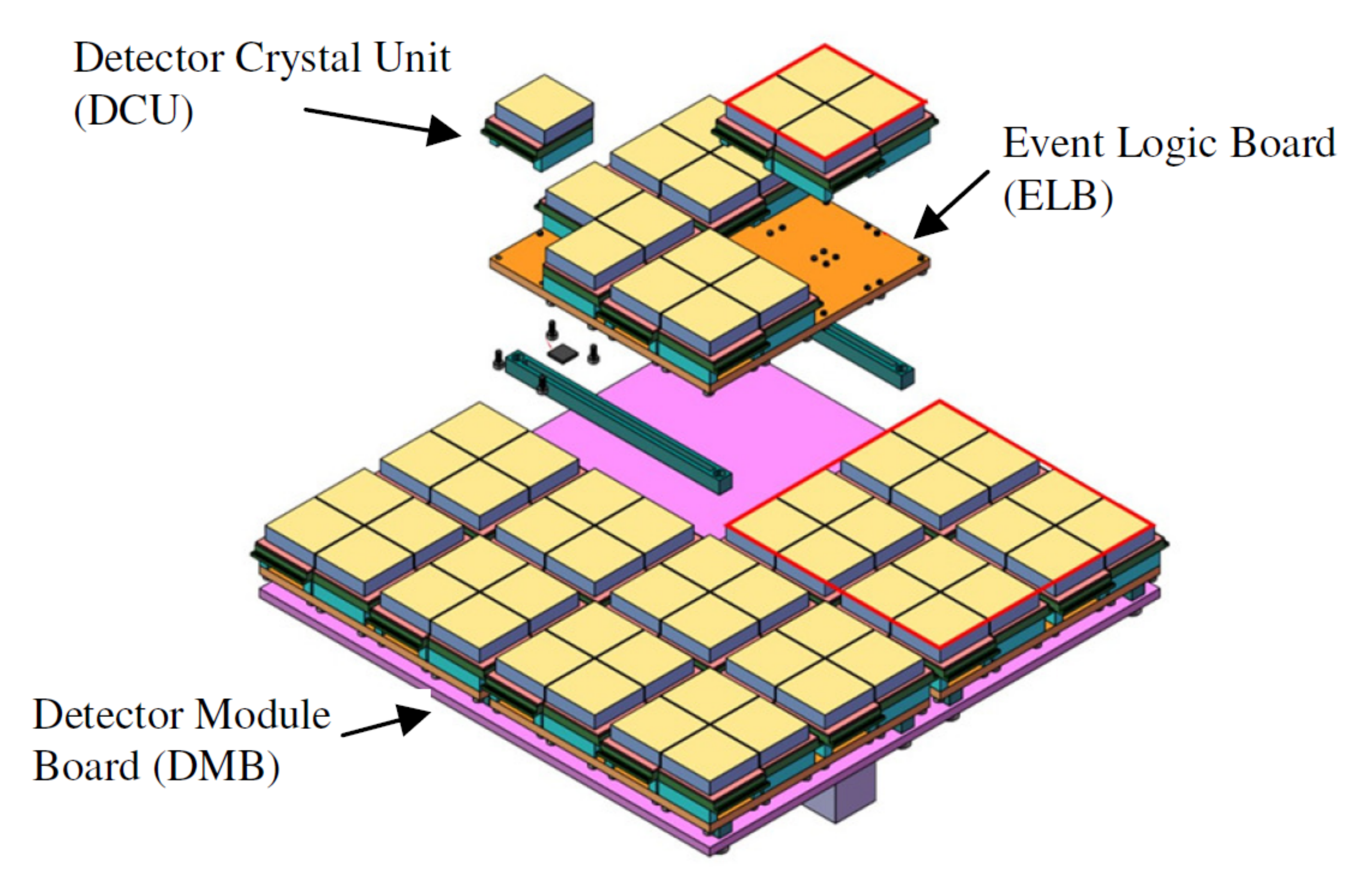}
    \caption{P2 architecture: P2 detector plane is composed of 64 Detector Crystal Units. A DCU consists of a hybridized 19.9\thinspace mm $\times$ 19.9\thinspace mm $\times$ 5 mm CZT detector, with 0.6 mm pixel pitch. A Detector Crystal Array (DCA) is formed by an array of $4\times4$ DCUs. Grouping 4 DCAs an Event Logic Board (ELB) is formed. A populated ELB is called Quad Detector Module (QDM). It contains 16 Complex Programmable Logic Devices (CPLD) responsible for the low logic interface to each of the NuASIC. At last, a closely tiled array of $2\times2$ ELBs is connected to a Detector Mother Board (DMB) which controls and processes the 64 DCUs through a Field Programmable Gate Array (FPGA).}
    \label{p2}
  \end{center}
\end{figure}

In order to demonstrate the performance for the next generation of hard X-ray surveys, the {\it ProtoEXIST} program has developed and tested 2 coded mask hard X-ray telescopes, P1 and P2. Both include a moderate area ($\sim 256$\thinspace cm$^{2}$), low power, low electronics noise, close-tiled array of pixelated 5\thinspace mm-thick CZT detectors, operating in the $5-200$\thinspace keV energy band. The individual detectors are manufactured by Redlen Technologies\footnote{http://www.redlen.ca}. The  {\it ProtoEXIST} basic detector plane is composed of 64 Detector Crystal Units (DCUs), arranged in $4\times4$ arrays of DCUs sitting on an Event Logic Board (ELB). An array of $2\times2$ ELBs form the Detector Mother Board (DMB), as described in Figure ~\ref{p2}. For P2, the detectors were re-metalized by Creative Electron Inc.\ (CEI)\footnote{http://creativeelectron.com} with a $600$\thinspace $\mu$m pixel pitch. P2 enhanced performance over P1  is achieved not only by the re-metalization of the detector surface with smaller-pitch contacts, but also by replacing the RadNet Aplication Specific Integrated Circuit (ASIC) used in P1 by the NuASIC ~\cite{Harrison10}, described below, which descends from the RadNet ASIC and was developed for the NuSTAR mission. Most importantly, on P2 the CZT detectors are directly bonded to the ASIC surface.

\subsection{The NuASIC}

\begin{figure}[!htb]
  \begin{center}
    \includegraphics[width=0.6\textwidth]{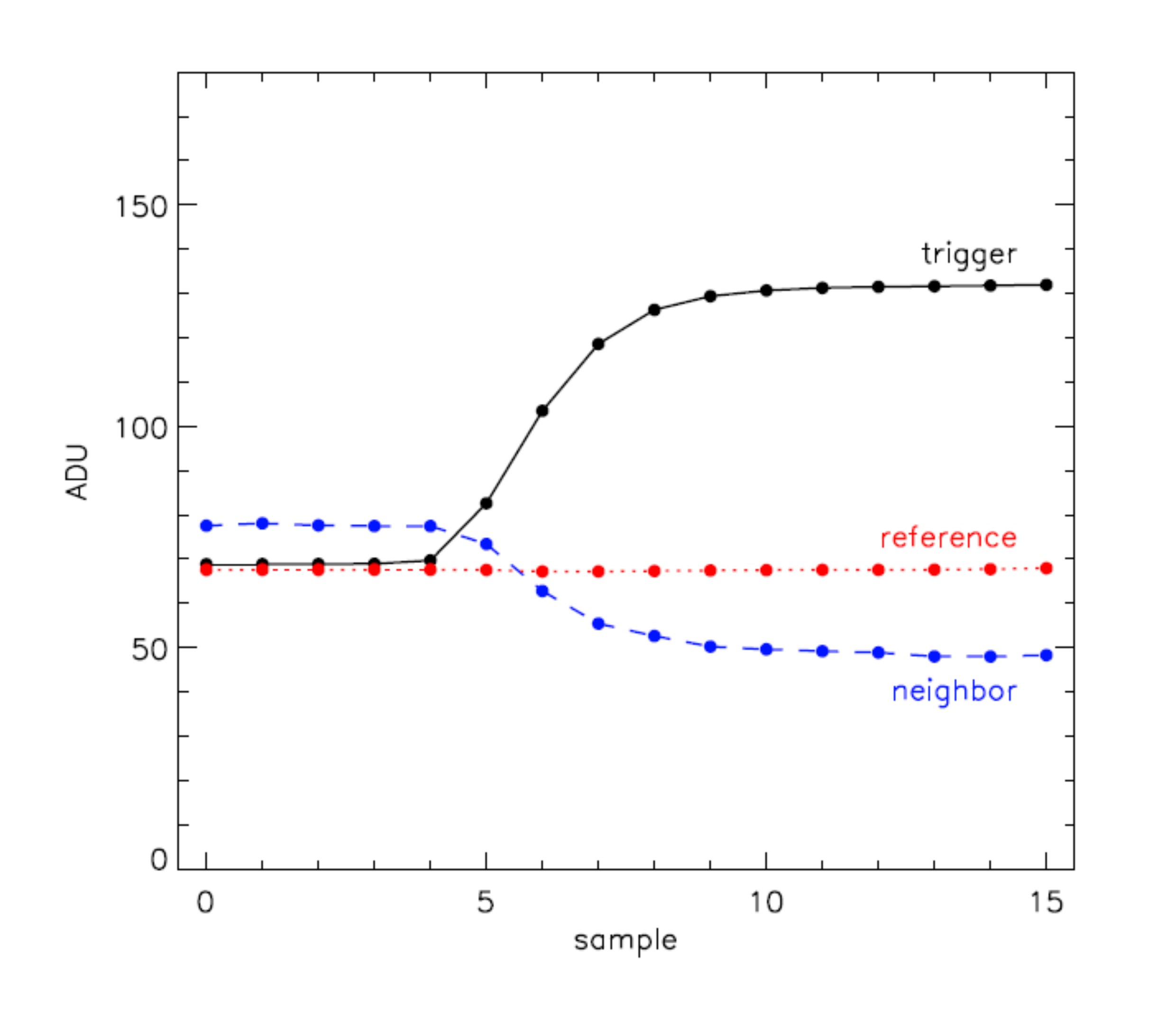}
\caption{Example pulse profiles averaged over events. For each event, the NuASIC can record 16 samples of the pulse profile for triggered pixels (solid black), neighboring pixels (dashed blue), and the rest of the pixels unrelated to the event (dotted red).}
  \label{pulse_profile}
 \end{center}
\end{figure}

The signal in the DCU is processed by the NuASIC, which contains 1024 channels (or pixels). Each pixel has a preamplifier which continuously amplifies input signals. The amplified signal is then sampled at a user-selected rate from 2 (normal) to 4 MHz. Then the output signal is sent to a bank of 16 capacitors for storage. Whenever a trigger occurs, the NuASIC samples and stores the 16 signals that represent pre-trigger samples, the pulse rise and post-triggers samples (see details in ~\cite{Kitagushi11}). Picture ~\ref{pulse_profile} illustrates an average over the pulse profiles of triggered pixels, as well as the neighbor pixels, and pixels unrelated to the event, sampled at 2 MHz. The pulse height information is obtained by subtracting the average of the pre-trigger from the average of the post-trigger samples. Averaging the pulse profiles over a large number of samples reduces the statistical flutuations in the electronic noise. The NuASIC preamplifiers and readout system are designed to allow for both electron and hole signals to be determined for each event, which enables the depth of interaction measurement. An internal test pulser (100 Hz) allows the injection of charge on individual pixels to determine the NuASIC electronic noise. In addition, the NuASIC can operate in the charge pump mode and in the normal mode. The charge pump mode, based on a charge pump circuit which provides a voltage that is higher than the voltage of the power supply or a voltage of reversed polarity, allows high efficiency, with eletronic noise of 400\thinspace eV (including leakage current) ~\cite{Harrison10}, but can handle leakage currents only up to 300 pA per pixel, constraining its operation under temperatures up to 10$^{\circ}$\thinspace C. In the normal mode, the NuASIC can be operated at room temperature, handling leakage currents up to 10\thinspace nA per pixel. In P2 tests and operation only the normal mode was used. 

Each NuASIC produced for the P2 DCUs is wirebonded to an ASIC Carrier Board (ACB), which in turn interfaces with the $32\times32$ CZT anode pixels in the detector surface. The output signal is digitalized on the ASIC to prevent extra capacitance noise. In order to evaluate the ACB functioning, the NuASIC is submitted to a test of the electronic noise, consisting on the application of the internal test pulser on a subset of pixels at a time. For this test, the NuASIC is placed on a test board (Figure ~\ref{test}), no high voltage bias is applied, and the internal pulser injects charges in a subset of $4\times4$ pixels for $40$s - long enough to generate a good Signal to Noise Ratio (SNR) to characterize the subset of pixels - stepping over the detector plane sequentially in the vertical axis, and in parallel on the horizontal axis. If validated, then the ACB is sent to hybridization. 

\subsection{CZT re-metalization and hybridization}

All Redlen CZT crystals acquired for the {\it ProtoEXIST} program already come with metalized anode and cathode, with gold pixels deposited on the anode side, in a configuration of $8\times8$ pixels with $2.5$\thinspace mm pixel pitch. In order to use the NuASIC in the P2 detectors, the Redlen CZT Crystals had to be submitted to the processes of re-metalization, which consists of polishing off the CZT surface for the complete removal of pixel contacts, and then the deposition of the new $32\times32$ pixel gold contacts. In addition, to suppress high leakage currents at the detector edges and a potential degradation of the edge pixels, a guard ring (GR) is placed on the anode side in the perimeter external to the edge pixels. This is done by CEI, which is also responsible for the hybridization process, consisting of bonding the CZT anode pixels to the corresponding pixels in the NuASIC using a conductive epoxy bond.

To improve the quality control of the CZT crystals in the process of hybridization, the re-metalized crystals are carefully chosen by inspection of their Infrared (IR) images and leakage current-voltage ($I\times V$) curves, prior to bonding. 

\section{DCU testing prior to P2 detector plane assembly}

A total of 65 DCUs were produced for the P2 October, 2012 balloon flight out of Fort Sumner, NM. Due to unexpected delays in the production of the ACBs, not all DCUs could be tested and approved in time for integration in the P2 detector plane assembly; 57 out of 64 could be integrated in time for the flight. The performance of each DCU was determined through a routine of tests during all stages of production, described as follows.

The tests to evaluate the detectors functionality and measure the noise contribution due to the bonding and leakage current are done in two steps. First, the DCUs are placed on the test board and the electronic noise is measured using the internal test pulser, as described previously. Secondly, the tests are repeated applying $-600$V on the surface of the CZT cathode with the aid of a conductive Al tape. To prevent shorts between the HV bias and the detector eletronics, an electrostatic side shield, made of Cu tape insulated using a Kapton tape, is placed around the edge of the CZT and connected to ground. The guard ring is kept at 0\thinspace V  and works as a sink for surface leakage currents on the edges of the detector. For details on the study of test pulser energy resolution of the NuASIC see ~\cite{Ballen11}. 

In order to prevent damage the test equipment in case of high voltage breakdown, characterized by the rapid increase in the leakage current up to values $\sim 2\times10^{3}$nA, the $I\times V$ curve of each DCU was obtained while slowly increasing the voltage applied to the CZT cathode. From a total of 50 DCUs that had their $I\times V$ curves measured prior to radiation tests, 3 of them had high voltage breakdown. The average leakage current on single DCU under a $-600$V bias is $\bar{I}=19.6\pm9.5$ nA (19 pA per pixel), with the most probable value of $10$ nA (9.8 pA per pixel).

\begin{figure}[!h]
 \centering
 \begin{tabular}{cc}
 \hspace{-0.1cm}\includegraphics[width=0.46\textwidth]{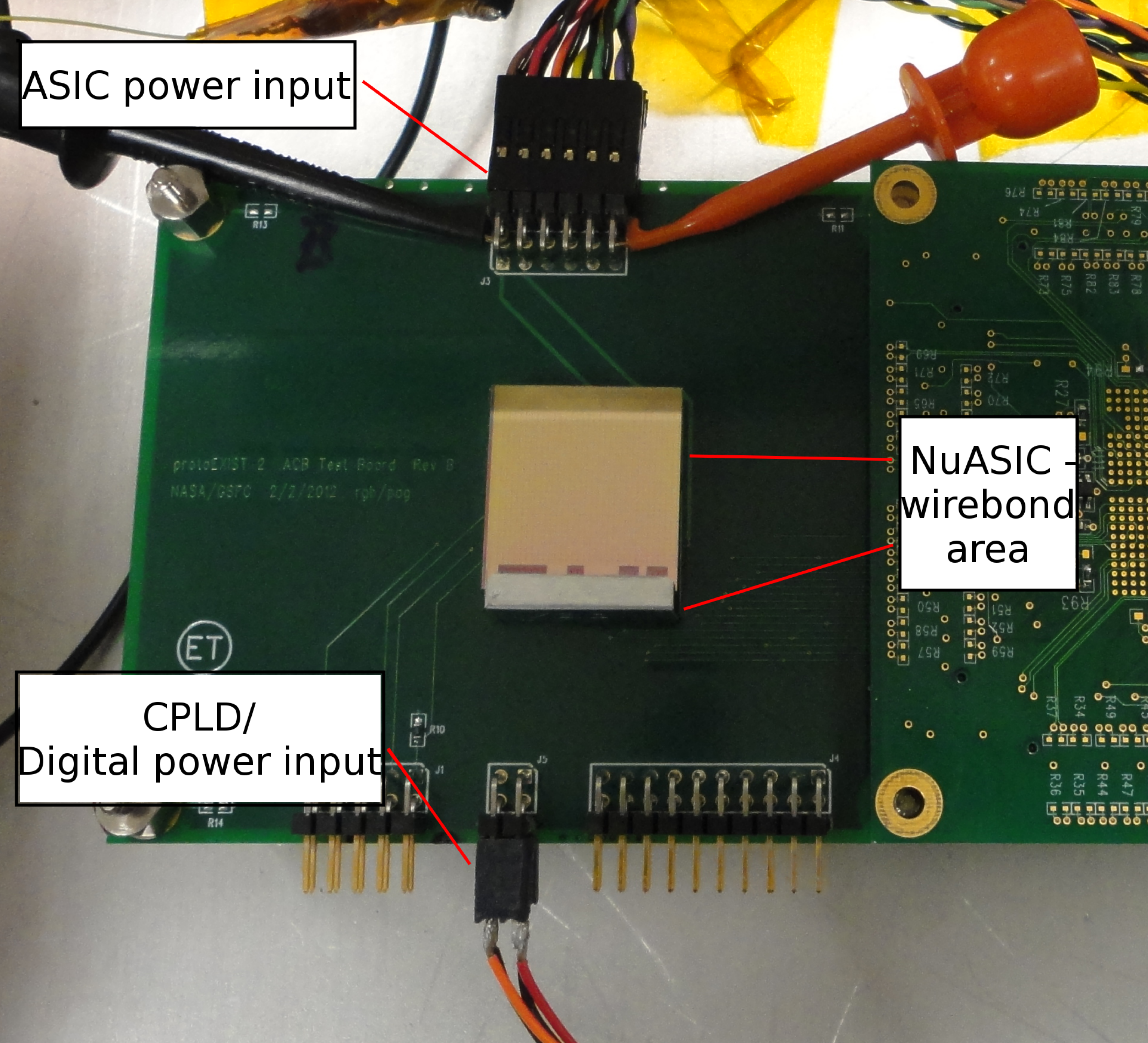} & \hspace{-0.05cm} \includegraphics[width=0.5\textwidth]{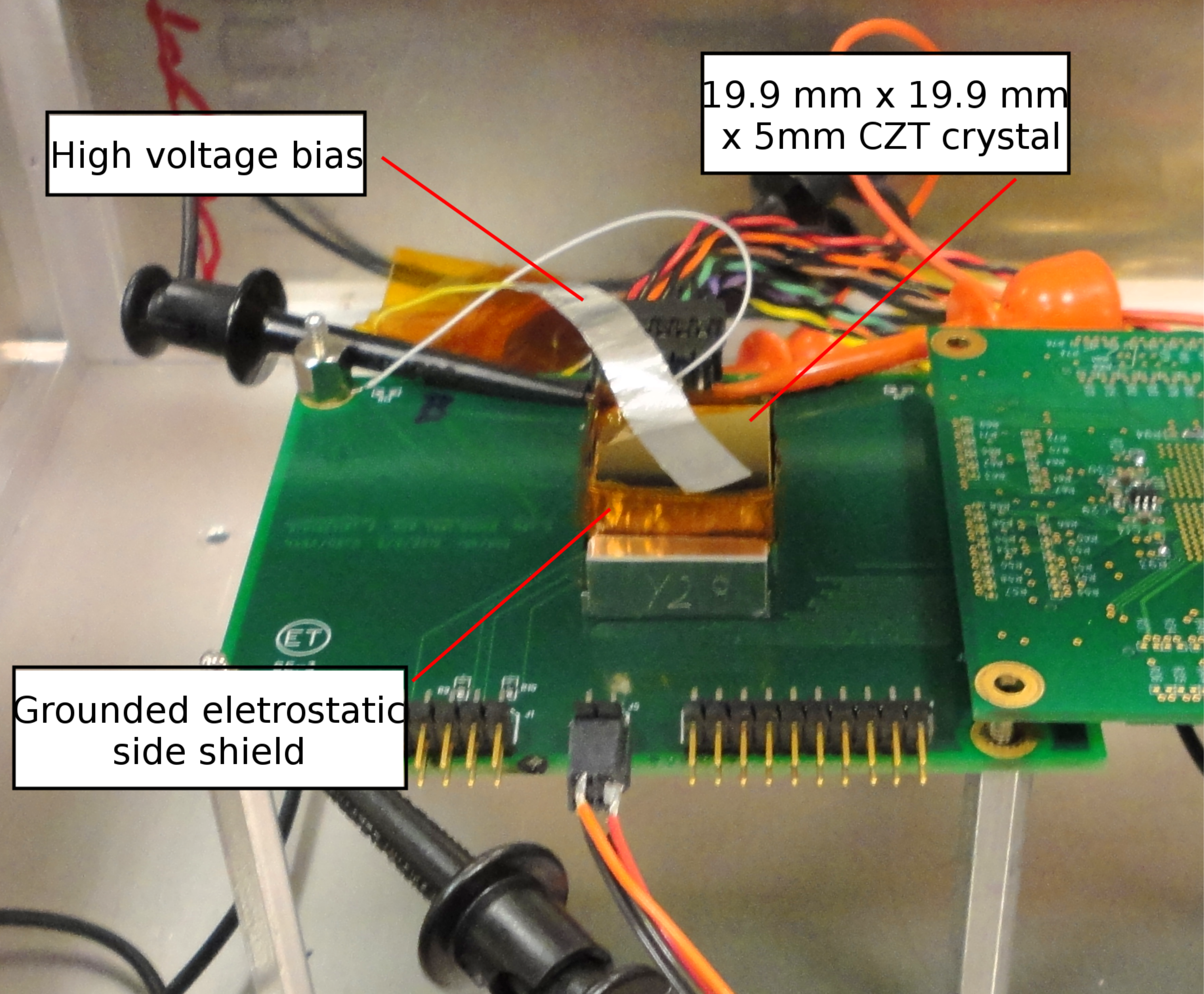} \\
 \end{tabular}
  \caption{A bare NuASIC sitting in the test board for evaluation of eletronic noise prior to hybridization (left). A DCU surrounded by the side shield placed on the test board (right). A conductive Al tape is connected to the cathode surface for the application of the high voltage bias.}
  \label{test}
 \end{figure}

Following the standard pulser tests, a radiation source - $^{241}$Am or $^{57}$Co - is placed above the test board to evaluate the DCU response in the energy band from 6 to 140 keV. The integration time for standard tests with a $10$mCi $^{241}$Am source is $2$h, whereas with a $200\,\mu$Ci $^{57}$Co source is $18$h. Triggered events are readout in a $5\times5$ region surrounding triggered pixels, in which: the triggered event is at the central pixel of the subarray; the $3\times3$ pixels surrounding the triggered event collect any charge which has leaked into neighboring pixels and also collect charges induced by holes, allowing the detection of split charge events and the reconstruction of the depth of interaction; the pixels at the perimeter of the $5\times5$ region allow the determination of the zero-point for later calibration. Figure ~\ref{test} shows the test board hosting a NuASIC and a DCU for standard tests.

\section{Data analysis and energy calibration of DCUs}

The reduction of the collected data is done by a pipeline that first selects only the events for which only one pixel was trigered, generating a single-trigger event file. Then, the energy response of the individual pixels is reconstructed. During the radiation data collection the triggered events are readout in an array of $5\times5$ pixels surrounding the triggered pixel. For single triggered events, the trigger occurs in the central pixel, and the readouts of the 16 pixels in the perimeter of the subarray register the ASIC response without the presence of an event. Thus, reading out the data from these pixels allows the determination of the zeropoint, or baseline, which gives the zero keV line for later energy calibration. 

The individual pixel response to an X-ray interaction, after amplification and shaping, is digitized and determines the pulse-height channel corresponding to the deposited energy. Due to the excellent linearity of the NuASIC response in the 5-200 keV energy range, the energy calibration solution of every pixel can be found by applying a simple linear regression to energy and channel values, taking into account the baseline at 0\thinspace keV. The energy channels are determined by the centroids of Gaussian fits to the photopeaks of the nuclear lines from the radioactive sources (59.6 and 122.0 keV). The angular coefficient of the linear equations found as the calibration solutions determine the gains of each pixel. Then, the energy resolution for each pixel can be calculated from the widths of the lines. 

\begin{figure}[!h]
 \centering
 \begin{tabular}{cc}
 \hspace{-0.5cm}\includegraphics[width=0.5\textwidth]{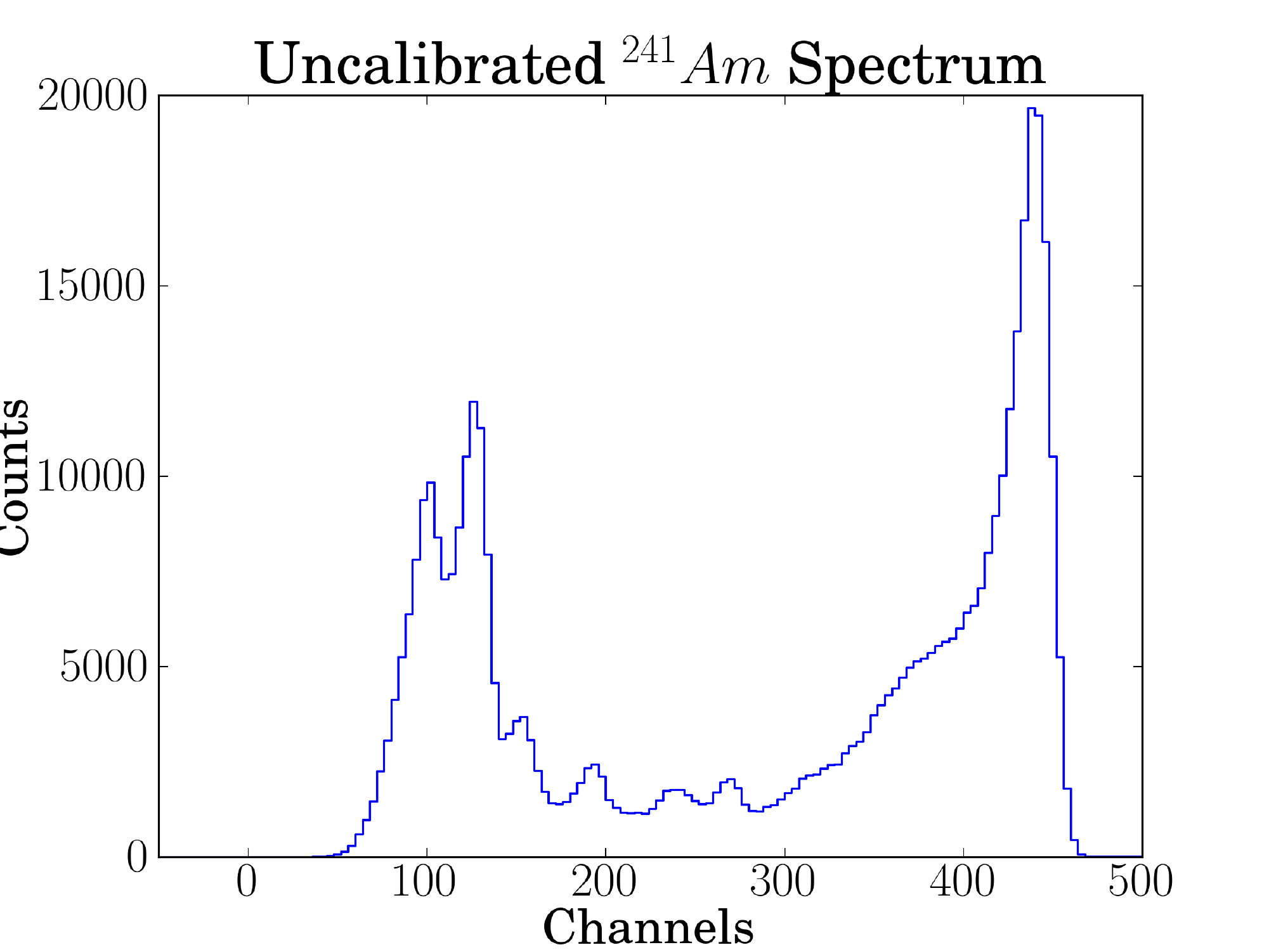} & \hspace{-0.5cm} \includegraphics[width=0.55\textwidth]{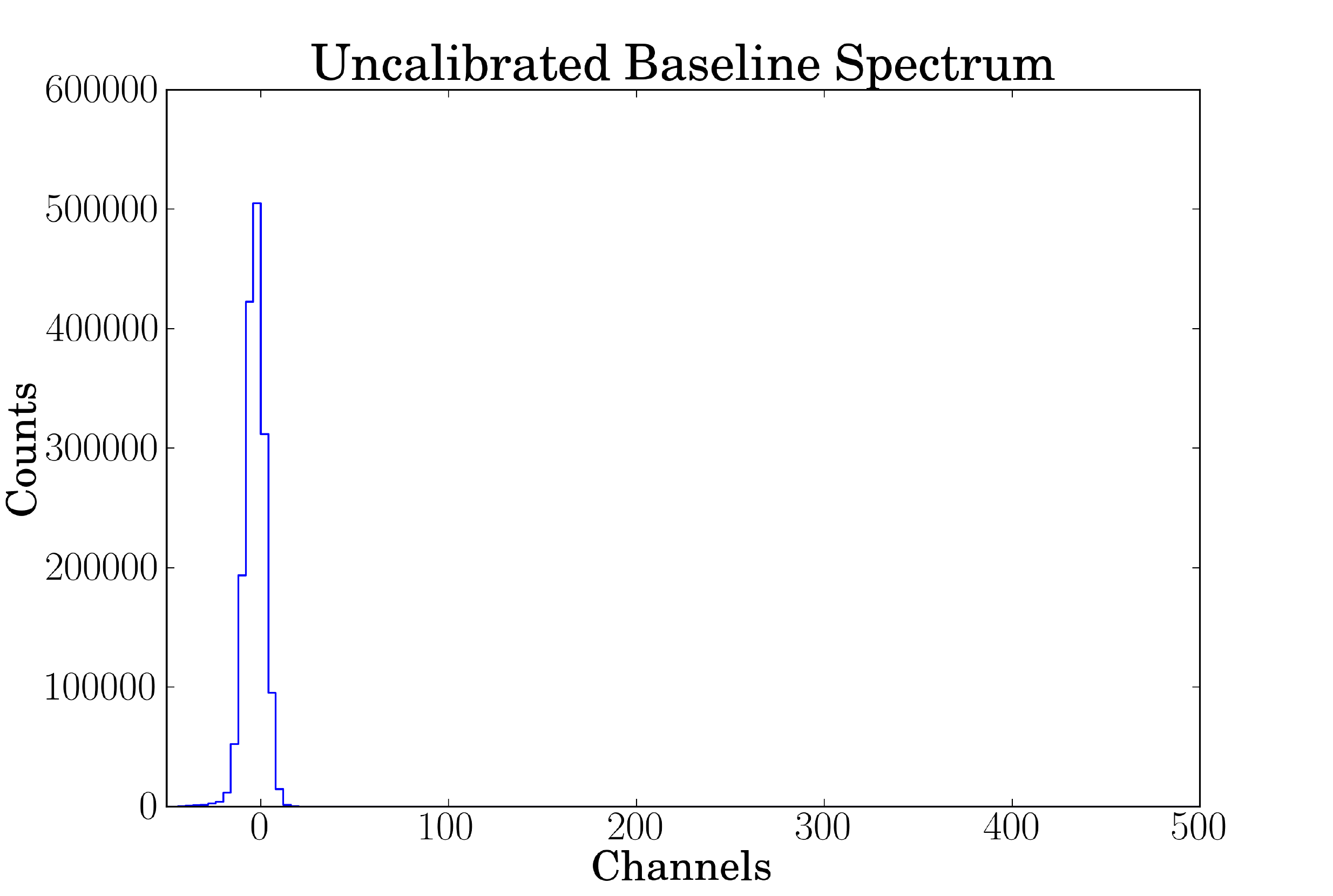} \\
 \hspace{-0.5cm}\includegraphics[width=0.5\textwidth]{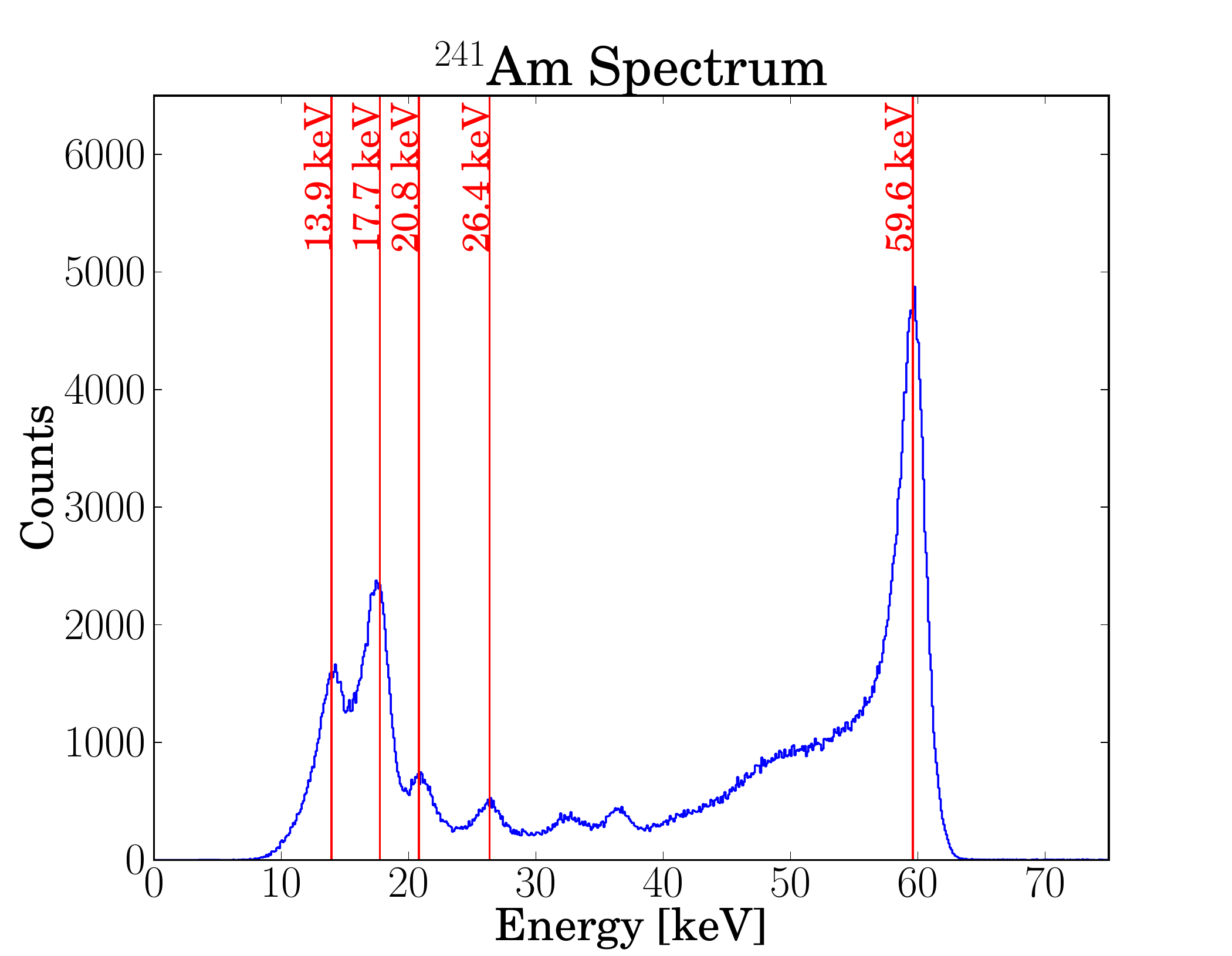} & \hspace{-0.5cm} \includegraphics[width=0.52\textwidth]{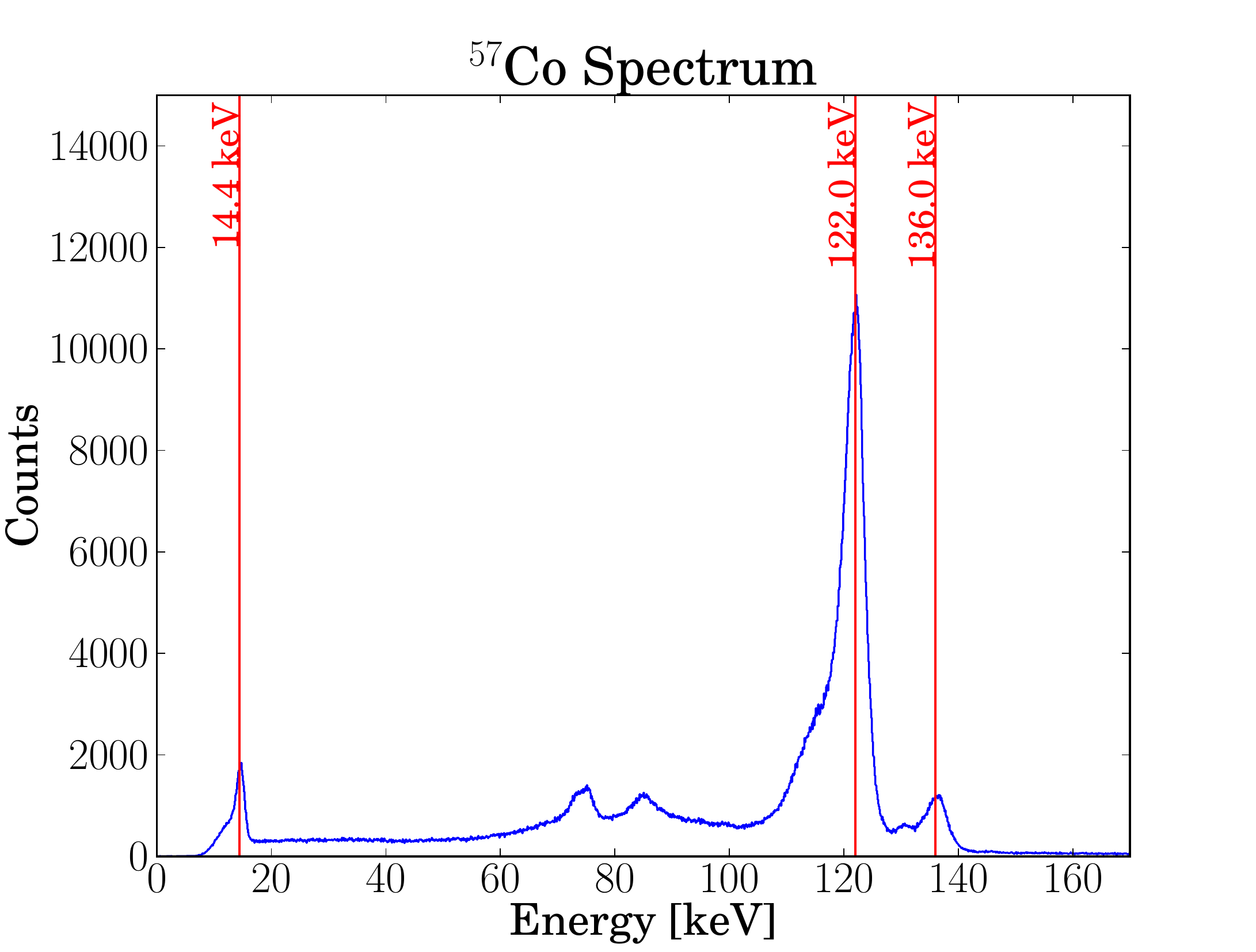} \\
 \end{tabular}
  \caption{Uncalibrated composite single trigger energy spectrum of an $^{241}$Am source (top left) and corresponding baseline (top right) obtained with DCU Y10. After the calibration, Y10 presents average energy resolution $\bar{E}_{FWHM}=1.98$ keV and average gain $\bar{G}=59.59$ ADC for the $60$keV photopeak. Bottom left: Composite single trigger energy spectrum of an $^{241}$Am source collected using DCU Y38 ($\bar{E}_{FWHM}=1.94$ keV). Bottom right: Composite single trigger energy spectrum of a $^{57}$Co source collected using DCU Z02 ($\bar{E}_{FWHM}=2.04$ keV).}
  \label{spec}
 \end{figure}

A composite single-trigger energy spectrum of an $^{241}$Am source and a baseline spectrum, both prior to energy calibration, as well as the energy calibrated single-trigger composite spectrum of a single DCU, are given in Figure ~\ref{spec}. Each DCU is identified by the NuASIC number corresponding to the letter code of the wafer on which the NuASIC was produced - X, Y, Z - followed by a number from $01$ to $49$.  

In order to improve the methods of searching the pulse profile within the overall radiation spectra, a maxima search smoothed over a 5-bin rolling averages to provide a better initial peak location was recently implemented to the fit tools. Furthermore, new tools to deal with photopeak asymmetries caused by incomplete charge collection were also implemented in our data reduction and analysis routine. As one can see in Figure ~\ref{spec}, this effect causes broadening in the energy resolution distribution which may lead to errors in the determination of the centroid of the protopeak and its FWHM. To avoid systematic errors in simple Gaussian fits, we have implemented a Gaussian plus a left tail function to fit the radiation profiles. The systematic fit uncertainties in resolution for Gaussian with a left tail fits for the individual pixels is $\sim 0.4$\thinspace keV (FWHM) for $^{241}$Am and $\sim 0.2$\thinspace keV (FWHM) for $^{57}$Co.
 
Table ~\ref{tableAm} and ~\ref{tableCo} shows median values of the energy resolutions (FWHM) obtained summing over all pixels in individual DCUs for the 59.6\thinspace keV and  122.0\thinspace keV, respectively. These results show a significant improvement in the energy resolutions over the ones obtained with 6 single prototype detectors developed for P2 during the first phase of development ($\lesssim4.1\%$ \@ 60 keV, and $\sim3.0\%$ \@ 122 keV ~\cite{Ballen11}), as well as a better performance than what was expected during P2 architecture planning ($\sim3.7\%$ \@ 60 keV, and $\sim3.0\%$ \@ 122 keV ~\cite{Hong09}).

\begin{table}[!htb]
 \centering
  \caption{FWHM median energy resolutions obtained summing over all pixels in individual DCUs for the \@ 59.6\thinspace keV line.}
   \begin{tabular}{cccc} \hline
    Detector ID & FWHM & $10^{th}$ percentile & $90^{th}$ percentile \\ \hline
    Y10 & 1.98 & 1.74 & 2.27\\
    Y23 & 2.07 & 1.70 & 2.61\\
    X19 & 2.14 & 1.86 & 2.58\\
    Y38 & 1.94 & 1.62 & 2.33\\
    Y06 & 2.17 & 1.75 & 2.68\\
    X10 & 1.95 & 1.56 & 2.39\\
    X40 & 2.00 & 1.63 & 2.49\\
    Y33 & 1.99 & 1.64 & 2.44\\
    X13 & 2.03 & 1.89 & 2.18\\
    X44 & 2.03 & 1.70 & 2.51\\
    X27 & 2.17 & 1.85 & 2.58\\
    X17 & 2.83 & 2.20 & 3.88\\
    Y32 & 2.15 & 1.77 & 2.62\\
    X25 & 1.98 & 1.60 & 2.56\\
    Z21 & 2.11 & 1.72 & 2.63\\
    Y27 & 1.98 & 1.49 & 2.77\\
    X23 & 2.14 & 1.65 & 2.86\\
    X24 & 2.14 & 1.72 & 2.67\\
    X45 & 2.25 & 1.82 & 2.63\\
    Z07 & 2.21 & 1.74 &3.00\\
    Y49 & 2.05 & 1.57 & 2.80\\
    Y16 & 2.06 & 1.55 & 2.68\\
    Z01 & 1.95 & 1.45 & 2.89\\
    Y21 & 2.37 & 1.79 & 3.14\\
    X14 & 2.07 & 1.83 & 2.58\\ \hline
   \end{tabular}
   \label{tableAm}
\end{table}

\begin{table}[!htb]
 \centering
  \caption{FWHM median energy resolutions obtained summing over all pixels in individual DCUs for the \@ 122.0\thinspace keV line.}
   \begin{tabular}{cccc} \hline
    Detector ID & FWHM & $10^{th}$ percentile & $90^{th}$ percentile \\ \hline
    X14 & 2.25 & 2.00 & 2.67\\
    X19 & 2.44 & 2.12 & 3.10\\
    X43 & 2.14 & 1.95 & 2.46\\
    Y02 & 2.55 & 2.24 & 3.22\\
    Y08 & 2.53 & 2.18 & 3.04\\
    Y18 & 2.16 & 1.83 & 2.71\\
    Y25 & 2.23 & 1.94 & 2.60\\
    Y37 & 2.13 & 2.02 & 2.21\\
    Y42 & 2.18 & 1.96 & 2.47\\
    Y46 & 2.17 & 1.96 & 2.42\\
    Y09 & 2.36 & 2.16 & 2.65\\
    Z02 & 2.04 & 1.86 & 2.25\\
    Z04 & 2.22 & 1.97 & 2.69\\
    Z05 & 2.13 & 1.86 & 2.59\\
    Z07 & 2.36 & 2.13 & 2.75\\
    Z15 & 2.16 & 1.95 & 2.53\\
    Z21 & 2.19 & 1.99 & 2.47\\ \hline
   \end{tabular}
   \label{tableCo}
\end{table}

\begin{figure}[!h]
  \begin{center}
    \hspace*{-1.5cm}\includegraphics[width=0.7\textwidth]{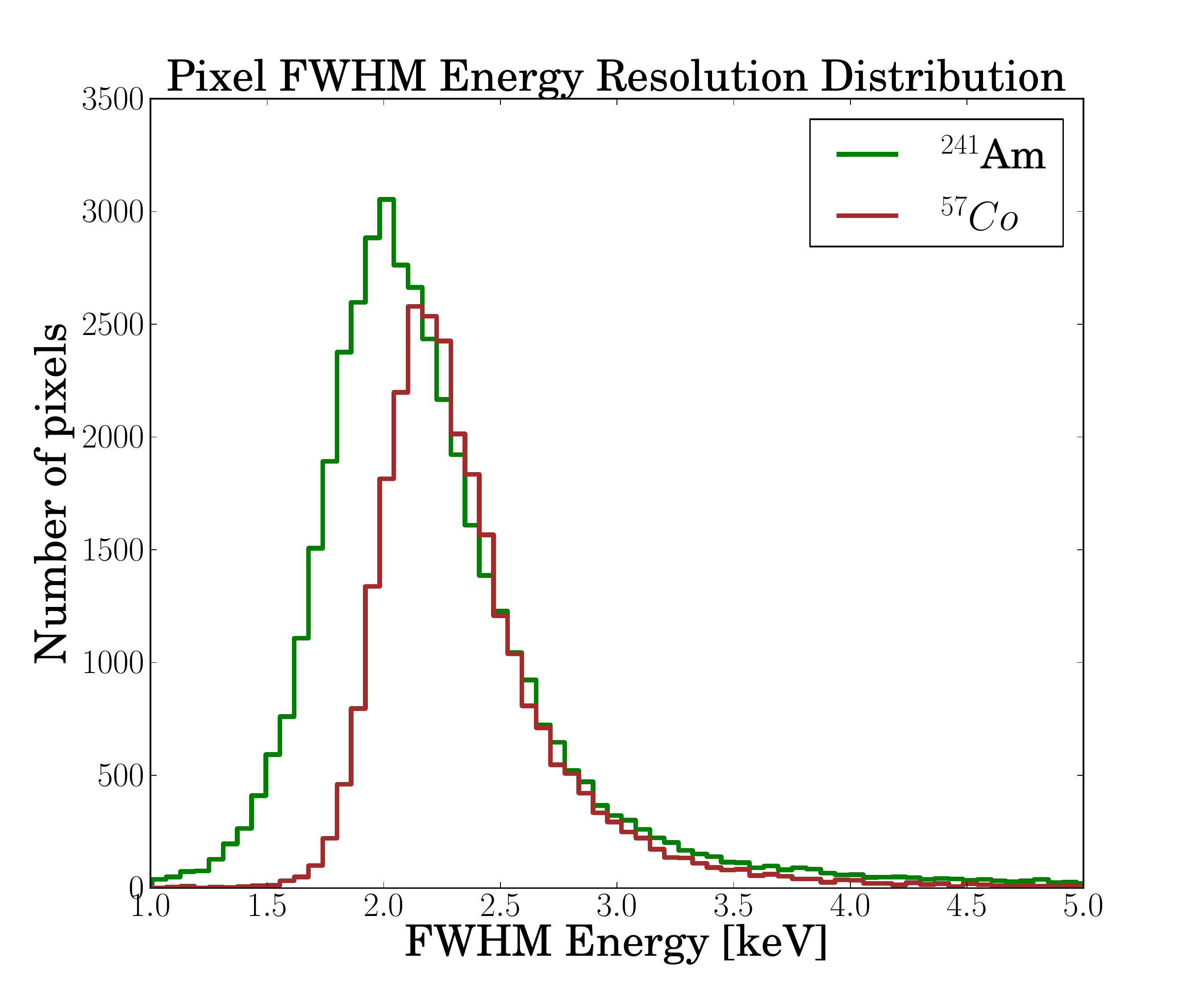}
    \caption{The FWHM energy resolution distribution of all pixels of 25 DCUs tested with $^{241}$Am (green) and 17 DCUs tested with $^{57}$Co (brown).}
    \label{distribution}
  \end{center}
\end{figure}

\begin{figure}[!h]
  \begin{center}
    \hspace{-0.8cm}\includegraphics[width=0.6\textwidth]{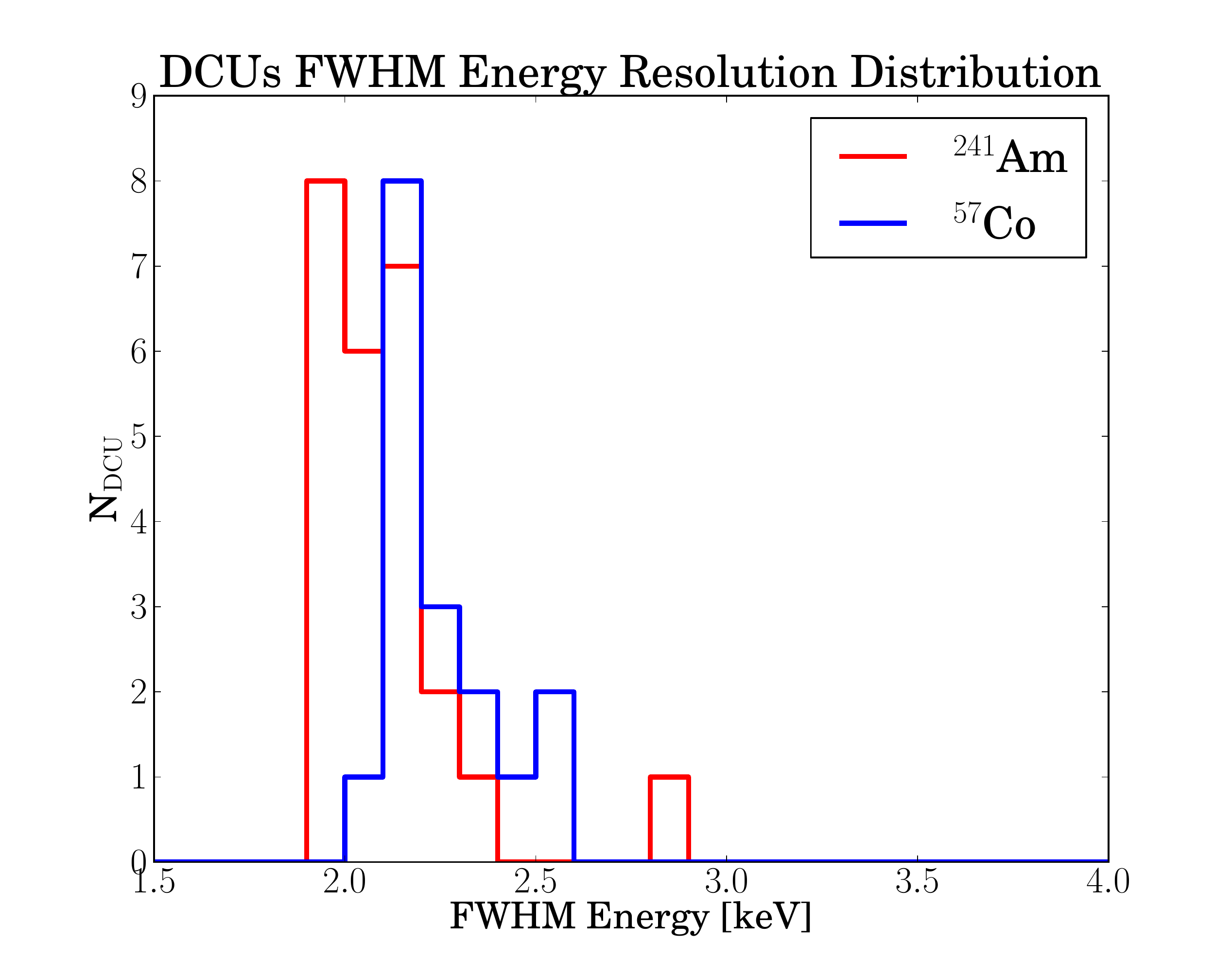} 
    \caption{The FWHM Energy distribution of $17$ DCUs tested with $^{57}$Co (blue) and of $25$ DCUs tested with $^{241}$Am (red). The median FWHM energy resolution obtained for the 122 keV line is 2.27 keV, whereas for the 60 keV line is 2.12 keV.}
    \label{resolution}
  \end{center}
\end{figure}

After the application of the calibration solution to individual DCUs, the single pixel energy resolution distribution is concentrated in the 1.0-4.0\thinspace keV range for the 60\thinspace keV and the 122\thinspace keV photopeaks. Figure ~\ref{distribution} shows this distribution for data collected using the $^{57}$Co source. The average values of the energy resolution of the 17 DCUs tested with $^{57}$Co is $2.27$\thinspace keV ($1.9\%$), and the average gain is 122.3\thinspace ADU, whereas the average value of the energy resolution of 25\thinspace DCUs tested with $^{241}$Am is 2.12\thinspace keV$(3.5\%)$, and the average gain is 59.7\thinspace ADU, as shown in Figure ~\ref{resolution}. Those energy resolution figures represent a significant improvement over P1 and demonstrates the good performance of the NuASIC. 

\section{Conclusions}

The {\it ProtoEXIST2} experiment achieved a significant improvement over {\it ProtoEXIST1\/} in both low-energy threshold ~\cite{Hong13} and energy resolution. The substitution of the RadNet ASIC used in P1 by the NuASIC in P2 enables an increase from 64 to 1024 pixels without changing the area of the CZT crystal unit, keeping the same efficiency in event readouts and allowing the energy resolutions to improve from $5.3\%$ to $3.5\%$ (2.12\thinspace keV FWHM) at $59.6$\thinspace keV, and from 3.1\% to 1.86\% (2.27\thinspace keV FWHM) at 122.0\thinspace keV. The low-energy threshold for P1 is in the 25 to 30\thinspace keV range, whereas for P2 a $\sim$6\thinspace keV threshold has already been obtained in the normal mode operation. 

For the current configuration of P2, we demonstrate in this study that it is possible to obtain energy resolutions as low as 1.94\thinspace keV FWHM at 59.6\thinspace keV and 2.04\thinspace keV at 122.0\thinspace keV for individual DCUs at room temperature.

The P2 detector plane has been successfully tested in a near space-environment during a high-altitude balloon flight. Preliminary results of the flight data ~\cite{Hong13} show a FWHM average energy resolution $\lesssim2.4$\thinspace keV at 59.6\thinspace keV.

The high energy resolution achieved by P2 in a wide energy band, as well as the high spatial resolution obtained with the P2 CZT detectors ($0.6$ mm pixel pitch), are very well suited for the MIRAX mission requirements. 

\acknowledgments

We would like to thank MIRAX and {\it ProtoEXIST} technical teams for the great effort done for this mission development. Special thanks to J. Grindlay and CAPES for fellowship grants. Development of the ProtoEXIST program and P2 detector and telescope at Harvard is currently supported by NASA grants NNX09AD76G and NNX11AF35G.

\end{document}